\documentclass[conference]{IEEEtran}

\makeatletter
\def\ps@IEEEtitlepagestyle{%
  \def\@oddfoot{\mycopyrightnotice}%
  \def\@evenfoot{}%
}
\def\mycopyrightnotice{%
  {\footnotesize XXX-X-XXXX-XXXX-X/XX/\$XX.00~\copyright~20XX IEEE\hfill}
  \gdef\mycopyrightnotice{}
}

\usepackage{blindtext}
\usepackage{eso-pic}
\IEEEoverridecommandlockouts
\usepackage{cite}
\usepackage{amsmath,amssymb,amsfonts}
\usepackage{algorithmic}
\usepackage{graphicx}
\usepackage{textcomp}
\usepackage{xcolor}
\def\BibTeX{{\rm B\kern-.05em{\sc i\kern-.025em b}\kern-.08em
    T\kern-.1667em\lower.7ex\hbox{E}\kern-.125emX}}
    
\usepackage{eso-pic}
\newcommand\AtPageUpperMyright[1]{\AtPageUpperLeft{%
 \put(\LenToUnit{0.17\paperwidth},\LenToUnit{-2cm}){%
     \parbox{0.9\textwidth}{\raggedleft\fontsize{8}{11}\selectfont #1}}%
 }}%
\newcommand{\conf}[1]{%
\AddToShipoutPictureBG*{%
\AtPageUpperMyright{#1}
}
}    
    
\begin{document}
\title{\vspace*{1cm} Self-Supervised Learning for Organs At Risk and Tumor Segmentation with Uncertainty Quantification
\thanks{
This project is supported by the NIH funding: R01-CA246704 and R01-CA240639, and FDOH (Florida Department of Health) through the James and Esther King Biomedical Research Program-20K04.}
}






\author{\IEEEauthorblockN{Ilkin Isler}
\IEEEauthorblockA{\textit{Department of Computer Science} \\
\textit{University of Central Florida}\\
Orlando, FL, USA \\
ilkin@knights.ucf.edu}
\and
\IEEEauthorblockN{Debesh Jha}
\IEEEauthorblockA{\textit{Department of Radiology} \\
\textit{Northwestern University}\\
Chicago, IL, USA \\
debesh.jha@northwestern.edu}
\and
\IEEEauthorblockN{Curtis Lisle}
\IEEEauthorblockA{\textit{KnowledgeVis} \\
\textit{LLC}\\
Altamonte Springs, FL, USA \\
clisle@knowledgevis.com}
\and
\IEEEauthorblockN{Justin Rineer}
\IEEEauthorblockA{\textit{Department of Radiation Oncology} \\
\textit{Orlando Health}\\
Orlando, FL, USA \\
justin.rineer@orlandohealth.com}
\and
\IEEEauthorblockN{Patrick Kelly}
\IEEEauthorblockA{\textit{Department of Radiation Oncology} \\
\textit{Orlando Health}\\
Orlando, FL, USA \\
patrick.kelly@orlandohealth.com}
\and
\IEEEauthorblockN{Bulent Aydogan}
\IEEEauthorblockA{\textit{Department of Radiation Oncology} \\
\textit{ University of Chicago}\\
Chicago, IL, USA \\
email address or ORCID}
\and
\IEEEauthorblockN{Mohamed Abazeed}
\IEEEauthorblockA{\textit{Department of Radiation Oncology} \\
\textit{Northwestern University}\\
Chicago, IL, USA \\
email address or ORCID}
\and
\IEEEauthorblockN{Damla Turgut}
\IEEEauthorblockA{\textit{Department of Computer Science} \\
\textit{University of Central Florida}\\
Orlando, FL, USA \\
damla.turgut@ucf.edu}
\and
\IEEEauthorblockN{Ulas Bagci}
\IEEEauthorblockA{\textit{Department of Radiology} \\
\textit{Northwestern University}\\
Chicago, IL, USA \\
ulas.bagci@northwestern.edu}
}

\maketitle
\conf{\textit{Proc. of the International Conference on Electrical, Computer, Communications and Mechatronics Engineering (ICECCME 2023) \\ 
19-20 July 2023, Tenerife, Canary Islands, Spain}}
\begin{abstract}
In this study, our goal is to show the impact of self-supervised pre-training of transformers for organ at risk (OAR) and tumor segmentation as compared to costly fully-supervised learning. The proposed algorithm is called Monte Carlo Transformer based U-Net (MC-Swin-U). Unlike many other available models, our approach presents uncertainty quantification with Monte Carlo dropout strategy while generating its voxel-wise prediction. We test and validate the proposed model on both public and one private datasets and evaluate  the gross tumor volume (GTV) as well as nearby risky organs' boundaries. We show that self-supervised pre-training approach improves the segmentation scores significantly while providing additional benefits for avoiding large-scale annotation costs.
\end{abstract}


\section{Introduction}

Radiation therapy (RT) is a common cancer treatment, in which radiation beams target cancerous tissues to destroy them. However, during this process, non-cancerous tissues and radiation-sensitive Organs At Risk (OARs) can be inadvertently exposed to damaging radiation. To minimize collateral damage, it is crucial to deliver radiation with precision, taking into account the size and location of the tumor. In this context, accurate segmentation of OARs and tumors plays a key role in effective radiation therapy planning.

Manual segmentation of organs and tumors is extremely time-consuming, expensive, and  prone to errors in boundary determination. Having inter- and intra-operator variability makes segmentation repeatability and standardization a challenging task. Despite these difficulties, manual delineation are still the reference paradigm in the current standards.


Deep learning models, particularly Fully Convolutional Neural Networks (CNNs) have shown great success for medical image segmentation tasks. The 'U' shaped architectures have been achieving state-of-the-art results for both 2D and 3D tumor and OAR segmentation across different tissue regions~\cite{ronneberger2015u, cciccek20163d}. However, available studies are limited to small cohort sizes with well-curated data sets, and more importantly, the state-of-the-art performances are not still at the clinically accepted levels. Given the importance of minimizing tissue toxicity and maximizing cancerous tissue targeting in treatment plans, accurate segmentation is more crucial than ever.


A notable limitation of deep learning based segmentation methods is that convolutional layers have a limited ability to capture long-range information which is essential for accurate OAR segmentation. In contrast, transformer models have recently demonstrated superior performance in capturing long-range information compared to previous architectures, including CNNs. In this regard, vision transformers (ViT)~\cite{dosovitskiy2020image}, for example, have been performing better than regular U-Net models because of their way of processing data as a sequence of patches and maintaining information about previous states. This approach allows transformers to better learn both local and global image representations. In medical image segmentation, understanding the relationships between organs, tumors, and their locations relative to each other is essential, as well as having a firm grasp of local features to ensure accurate pixel-wise classification.


\begin{figure*} [!t]
    \centering
    \includegraphics[width=17cm]{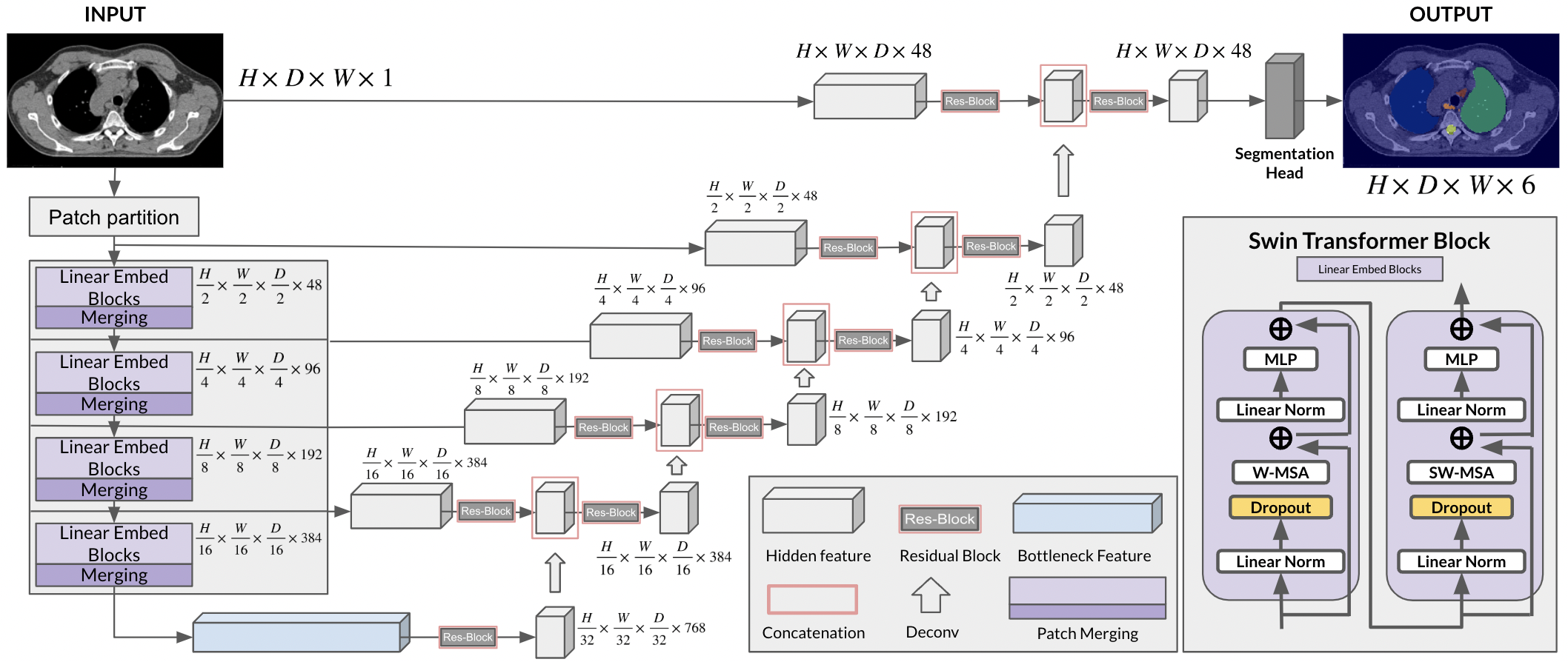}
    \caption{The proposed MC-Swin-U is illustrated. The architecture includes Monte Carlo dropouts to predict uncertainties, and Swin UNETR~\cite{hatamizadeh2022swin} to perform delineation.}
    \label{fig:architecture}
\end{figure*}

\section{Related Work}

Deep learning has brought about a major shift in the image segmentation domain, with convolutional neural networks (CNNs) emerging as the top method for numerous medical image segmentation tasks. The simple design and excellent performance of the U-shaped structure have led to numerous U-Net-like methods for both natural and medical images. Some notable examples include Res-UNet~\cite{Xiao2018WeightedRF}, U-Net++~\cite{zhou2018unet++}, UNet3+~\cite{huang2020unet}, 3D-Unet~\cite{cciccek20163d}, and V-Net~\cite{laina2016deeper}.

Among these techniques, some methods leverage neural architecture search to enhance segmentation performance. The nnU-Net framework~\cite{isensee2021nnu} has demonstrated impressive results by automating the design and training of neural networks for various medical image segmentation tasks, selecting the optimal network for a given dataset. Similarly, in their study, Guo et al., (2020)~\cite{guo2020organ} choose the most effective architecture for the specific organ involved. This paper utilizes stratified learning to improve the segmentation of organs at risk in head and neck cancer patients, leading to more accurate segmentation. Nonetheless, the proposed transformer-based methods have demonstrated their ability to achieve competitive results using a single architecture.

Trans-U-Net was one of the first papers to show having transformer layers in the model's encoder improves results~\cite{chen2021transunet}. UNETR was the first proposed architecture to be used within ViTs' encoder~\cite{hatamizadeh2022unetr}. Transformer-based U-Net architecture, has demonstrated strong capabilities in handling volumetric data and effectively segmenting organs at risk and tumors in various medical image datasets. Wang et al., (2021)~\cite{wang2021transbts} proposed TransBTS, a multimodal brain tumor segmentation model that uses transformers for feature extraction and classification. After the proposal of hierarchical vision transformer~\cite{liu2021swin}, where shifting windows are used to compute the self-attention, Swin UNETR has been proposed and shown to have great performance on the segmentation of brain tumors~\cite{hatamizadeh2022swin}. This architecture uses the Swin transformers in the encoder and is connected to a CNN-based decoder with skip connections. 3D images are reduced to a 1D sequence of embedding to be given as input to the Swin transformers. 

Although vision transformer-based methods exhibit strong performance, most ViT models achieve more robust intermediate representations when pre-trained on large datasets~\cite{raghu2021vision} and medical imaging cannot maximally benefit from the most common pre-training datasets since they generally consist of high resolution detailed data unlike medical scans. Pre-training on large medical segmentation datasets is also challenging due to the scarcity of well-annotated medical data. Even when such datasets are available, it is difficult to rely on the accuracy of the labels due to high intra- and inter-operator variability. Self-supervised learning and the proxy tasks chosen for it have proven effective in teaching models the texture, anatomical structures, and structural content of images. The self-supervised pre-training of Swin transformers has been studied and demonstrated to be effective when used on downstream tasks after fine-tuning the model on the target data~\cite{tang2022self}. In light of these factors, we have chosen to utilize a self-supervised pre-trained Swin transformer for our study.


Due to the high-risk nature of radiation therapy planning, concerns about the trustability of using artificial intelligence are a reality. In this regard, delivering all the possibilities and segmentation uncertainties into the evaluation is important. To this end, uncertainty quantification of the segmentation maps can be presented to the oncologists and technologists for a complete evaluation framework. Uncertainty information can be obtained in different ways. Deterministic networks, ensemble methods, and probabilistic methods are some of the mostly used approaches for obtaining uncertainty information. Test-time augmentations can be an alternative option to generate multiple outputs from one single deterministic network. Lastly, there are several different stochastic deep neural networks using Bayesian approaches, where repeated forward passes will generate different predictions~\cite{gawlikowski2021survey}. In this work, we'll be using  Monte Carlo Dropout (MCDO)~\cite{gal2016dropout} method to estimate the predictive uncertainty by enabling Dropout layers at test time. This will provide us/users to validate the segmentation results from boundary uncertainties when necessary.

In this work, we propose a Monte Carlo Transformer-based U-Net (MC-Swin-U) architecture for tumor and OAR segmentation in radiation oncology settings. Our approach combines the strengths of self-supervised learning, Swin Transformers, and uncertainty estimation to deliver improved segmentation results and provide clinicians with valuable information about the uncertainty associated with model predictions.


The key contributions of this paper can be summarized as follows:
\begin{enumerate}
\item Segmentation accuracy is increased by utilizing self-supervised pre-trained weights.
\item The requirement for large-scale, well-curated, and labeled data in deep learning-based segmentation training is avoided through self-supervised learning.
\item Various segmentation possibilities are better understood by oncologists when prediction results are coupled with uncertainty quantification. This mimics the decision-making process where multiple radiologists collaborate to perform segmentation together.
\item The effectiveness of the proposed framework is validated by fine-tuning pre-trained Swin Transformers on both public (NSCLC) and private (Orlando Health) benchmarks. In both cases, improvements in organ at risk and tumor segmentation results are observed due to pre-training.
\end{enumerate}


Our study is innovative as we are not aware of any study utilizing self-supervised learning with  uncertainty estimation as applied to both OAR and tumors segmentation in radiation oncology settings.

\section{The MC-Swin-U architecture}

Figure~\ref{fig:architecture} displays the block diagram of the proposed Monte Carlo Transformer based U-Net (MC-Swin-U) architecture. This network is referred to as MC-Swin-U network because we generate uncertainty in predictions through test-time dropout to approximate the uncertainty of a Bayesian network at a fraction of the computational cost, and we use Swin UNETR as our segmentation engine~\cite{https://doi.org/10.48550/arxiv.1910.10793}.

The U-Net architecture, created for biomedical image segmentation, has a U-shape with an encoder and decoder connected by skip connections. The encoder extracts high-level features through convolutional and downsampling layers, while the decoder upsamples the features and combines them with high-resolution encoder features via skip connections. This helps retain spatial information and generate precise pixel-level segmentation predictions.

In our MC-Swin-U architecture, we adopt the Swin UNETR as the underlying segmentation engine, which incorporates the strengths of Swin Transformer blocks into a U-Net-like architecture. This allows us to leverage the power of Transformers in image segmentation tasks. The MC-Swin-U architecture is self-explanatory, with details of the filter types, numbers, size, and other architectural design illustrated in Figure~\ref{fig:architecture}.

\begin{table} [t!]
\centering
\caption{Ablation Study of different architectures on the Orlando Health dataset with conventional supervised learning. DICE and Hausdorff Distance 95 scores are shown.}
\label{tab:ablation1}
\begin{tabular}{|l|l|l|}
\cline{1-3}
 OH & \textbf{Dice} & \textbf{HD95 (mm)}\\
 \hline
UNet-3D~\cite{ronneberger2015u}          & 0.743   & 46.873   \\ 
\hline
DynUNet~\cite{isensee2021nnu}          & 0.759   & \textbf{34.192}    \\ 
\hline
UNETR~\cite{hatamizadeh2022unetr}     & 0.750  & 37.766    \\ 
\hline
SwinUNETR~\cite{hatamizadeh2022swin}     & \textbf{0.761}  & 36.131   \\ 
\hline
\end{tabular}
\end{table}

\begin{table} [t!]
\centering
\caption{Ablation Study of the proposed MCSwinU. DICE scores are shown in the table. Here, 'PT', 'w', and 'w/o' refers to pre-training, with and without respectively.}
\label{tab:ablation2}
\begin{tabular}{|l|l|l|l|}
\cline{1-4}
 OH & \textbf{SwinUNETR} & \textbf{MCSwinU w/o PT}& \textbf{MCSwinU w PT}\\
 \hline
Lung R          & 0.960   & 0.971 & 0.973   \\ 
\hline
Lung L         & 0.957   & 0.968 & 0.971    \\ 
\hline
Spinal Cord     & 0.841  & 0.858 & 0.850    \\ 
\hline
Esophagus     & 0.573  & 0.576 & 0.589   \\ 
\hline
GTV     & 0.438  & 0.449 & 0.486   \\ 
\hline
\textbf{Overall}     & 0.761  & 0.764 & \textbf{0.773}   \\ 
\hline
\end{tabular}
\end{table}

An ideal architecture should have self-supervised learning capabilities, effective attention mechanisms for better object boundary localization, and the ability to handle multi-resolution details. The Self-Supervised SwinUNet-R architecture meets these criteria and has been identified as one of the state-of-the-art models~\cite{tang2022self}. To evaluate its suitability for our tumor and OAR segmentation problem, we conducted an ablation study and compared the performance of different architecture variants. The results, presented in Table~\ref{tab:ablation1}, show that SwinUNet-R achieved the highest dice score among other variants. Therefore, we selected SwinUNet-R as the backbone architecture for our subsequent experiments. The results presented in Table~\ref{tab:ablation2} indicate that adding dropout layers to SwinUNet-R increased the DICE score, as shown in the second column labeled MCSwinU w/o PT. Additionally, incorporating pre-training on top of the SwinUNet-R architecture with dropout layers (MCSwinU w/o PT) resulted in a 1\% increase in the overall DICE score, as demonstrated in the column labeled MCSwinU w PT. 

MC-Swin-U consists of two major parts: a Swin Transformer encoder and a CNN-based decoder that are connected to each other via skip connections at different resolutions. The input to our model is 3D (volumetric) CT images. The patch partition module generates sequences of 3D tokens that are reduced to a 1D sequence of embedding to be given as input to the Swin Transformer Blocks. The Swin Transformer Blocks contain dropout, multi-layer perceptron (MLP), linear normalization, regular (W-MSA), and window partitioning (SW-MSA) multi-head self-attention modules. From there, the extracted feature representations are then given to the CNN-based decoder that has a convolutional layer and a sigmoid activation function at the end to output the 6-channel (5 classes + background) prediction.

The goal of our study is to show the impact of using self-supervised pre-training compared to conventional supervised learning. To achieve this, we selected a pre-trained Swin transformer using weights from training on over 5000 volumetric CT images using self-supervised training methods, including Head \& Neck Squamous Cell Carcinoma, Lung Nodule Analysis 2016, TCIA CT Colonography Trial, TCIA Covid 19 and TCIA LIDC datasets~\cite{tang2022self}. In self-supervision, the key is to generate pairwise data: for this purpose, sub-volumes from the 3D images were augmented with rotation and cut out and used for pre-text tasks which are inpainting, contrastive learning, and rotation. We then fine-tuned our proposed network with this pre-trained (with self-supervision) model on a downstream task, which is tumor and OAR segmentation using labeled data in our application. To be more specific, the existing pre-trained weights were fine-tuned according to the model on our target dataset for 40,000 iterations. We validated the model performance for every 500th iteration. We used an image patch size of 96 $\times$ 96 $\times$ 96, a feature set depth of 48, and set the dropout rate to 0.5. We compared our self-supervised strategy with conventional supervised strategy. 

In summary, the proposed MC-Swin-U architecture leverages the strengths of the self-supervised SwinUNet-R model, which fulfills the requirements for effective object boundary localization and handling multi-resolution details in tumor and OAR segmentation problems. By incorporating fine-tuning on our target data, our approach demonstrates improved performance compared to conventional supervised learning methods.

\textbf{Uncertainty quantification:} Uncertainty is an integral outcome of our design, serving as a guidance for clinicians regarding segmentation evaluations, calibration, and accuracy of the results. In other words, the main reason behind adding uncertainty into our model is to highlight the areas in the segmentation results to help radiation oncologists for visual and quantitative evaluation, and give confidence visualization about the boundaries where the most uncertain regions concur with less confident regions. This aspect can be crucial for treatment planning. Utilizing uncertainty for segmentation guidance will be explored in future work.

The uncertainty quantification for each prediction is derived from Monte Carlo samples after training the deep learning model. During testing, we set the dropout layers to train mode, ensuring their activation during forward inference. We assessed the model 10 times and designated areas as 'uncertain' if fewer than 5 predictions agreed on that specific area (pixel).

\begin{figure*}
    \centering
    \includegraphics[width=18cm]{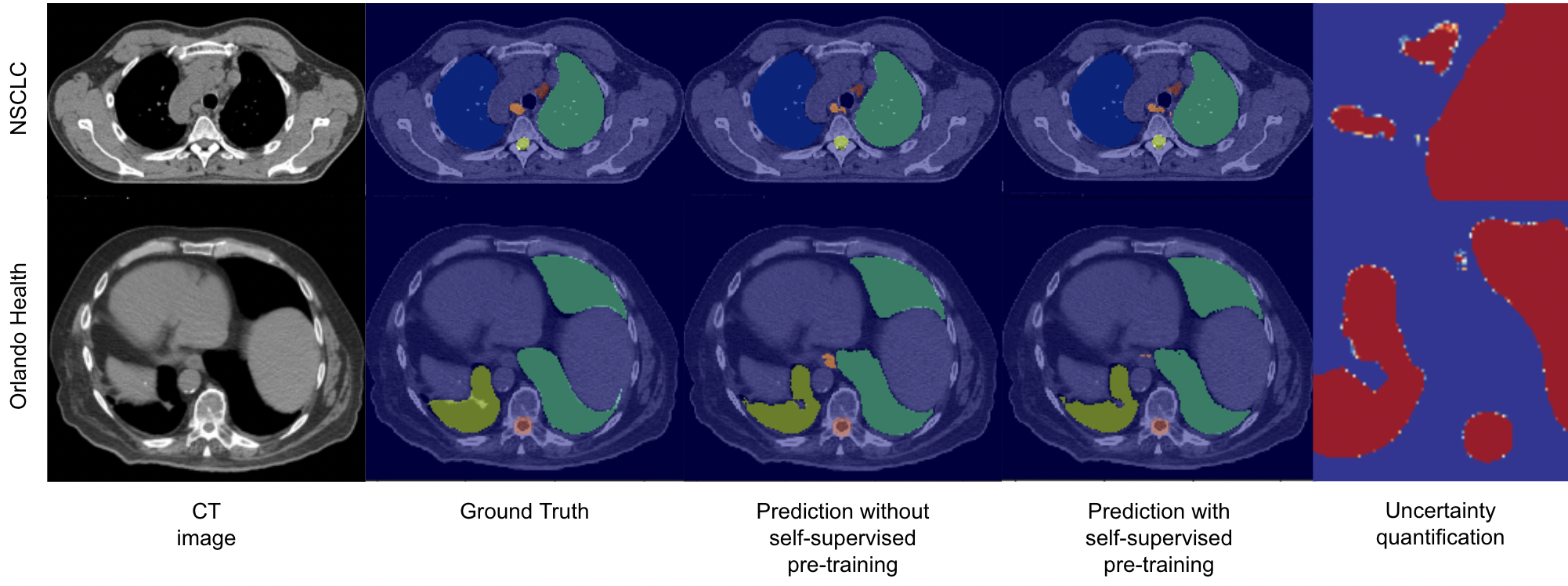}
  \caption{NSCLC (open) and Orlando Health (private) data are shown for OAR and tumor segmentation. The second column is ground truth, the third column is the resulting segmentation without self-supervised pre-training while the fourth column demonstrates the improved results with self-supervised pre-training. The last column presents zoomed uncertainty maps, mostly localized on the boundaries as expected.}
    \label{fig:results}
\end{figure*}
\section{Experiments}
\subsection{Dataset}
We demonstrated the effectiveness of our methods using one publicly available dataset and one private dataset provided by clinicians from Orlando Health. We have pre-processed both datasets in a way that each has the same organ structures delineated: right lung, left lung, spinal cord, and esophagus for OAR, and gross tumor volume (GTV) for lung tumor(s). The organs with larger regions would be the easiest to segment which would be the lunges in our case. The organs with a lower volume/surface area ratio (longer shape and a lower volume), can be categorized as moderately difficult to segment. Hard-to-segment organs would be those  with the smallest volume, such as the esophagus. Finally, any lung  tumors can be considered as the most difficult segmentation target because the size, location, and shape can be completely different for each patient, and even varying over time for the same subject. 

\textbf{NSCLC-Radiomics}~\cite{aerts2014decoding}: This collection contains thoracic region images of non-small cell lung cancer (NSCLC) patients  and their manual delineations (primary gross tumor volume ("GTV-1") and selected anatomical structures (i.e., lung, heart and esophagus)). After extracting DICOM imagery from The Cancer Imaging Archive (TCIA)~\cite{clark2013cancer}, we chose a subset of 133 patients who have ground truth structures for all 4 organs (right lung, left lung, spinal cord, esophagus) and the GTV.

\textbf{Orlando Health}: This is a private dataset from a local partner hospital, Orlando Health. This dataset consists of 3D CT images and manually delineated structures for over 250 NSCLC patients treated using Stereotactic body radiotherapy (SBRT) between 2011 and 2018.  We selected 144 patients out of this dataset that had ground truth contours for the same structures as our NSCLC dataset. The CT images were captured with a voxel size  of 1.7mm x 1.7mm in-plane and 3mm in the out-of-plane direction. 
\subsection{Implementation details}
We have implemented the proposed architecture using the PyTorch~\cite{paszke2019pytorch} framework. The experiment were run on NVIDIA GeForce RTX 3090 GPU. Both datasets used in the study had a resolution of 512 $\times$ 512 $\times$ Z, where Z varies according to the scan. Both were reformatted into NumPy tensors for deep learning at native resolution without resampling. For data augmentation, we applied well-known random transforms, using the emerging MONAI medical imaging framework~\cite{MONAI_Consortium_MONAI_Medical_Open_2020}: including intensity scaling and shifting, boundary cropping, affine transformation, 3D elastic transformation, and random Gaussian noise. We partitioned the data sets into train, validation, and test with ratios of 0.7, 0.15, 0.15, respectively. We present the test results in Table~\ref{tab:results}. 

\vspace{-1mm}
\section{Results and discussion}
We utilized two different conventional metrics to evaluate our approach: DICE and Hausdorff Distance 95 (HD95). The DICE coefficient is broadly used in medical image segmentation, indicating the region-wise similarities (overlaps) between two segmented objects. Hausdorff distance is a metric for calculating the dissimilarity between the boundaries of two segmented objects. The reason behind using the 95\% is to eliminate the impact 5\% of the outliers.

Having the same OAR ground truth for both datasets allowed us to compare the segmentation predictions from our network across the datasets. We observed the performance of the MC-Swin-U when using self-supervised pre-training on a public dataset results in consistent outcomes across both public and private datasets. This was an important point to show in a clinical setting, radiation oncologists can employ these pre-trained transformer networks to improve their segmentation results without having to perform costly training of a network from scratch on their own imagery.

\begin{table} [t!]
\centering
\caption{Evaluating the effect of self-supervised pre-training on the public NSCLC dataset and the private Orlando Health with MC-Swin-U for 40,000 iterations.}
\label{tab:results}
\begin{tabular}{|l|l|l|l|l|}
\hline
   & \multicolumn{2}{l|}{\textbf{Without SS pre-training}} & \multicolumn{2}{l|}{\textbf{With SS pre-training}}  
\\ 
\cline{2-5}
 NSCLC & \textbf{Dice} & \textbf{HD95 (mm)} & \textbf{Dice} & \textbf{HD95 (mm) }\\
 \hline
Lung R          & 0.983   & 18.849  & 0.983  & 2.611  \\ 
\hline
Lung L          & 0.981   & 36.491  & 0.982  & 2.066  \\ 
\hline
Spinal Cord     & 0.866  & 6.121  & 0.871  & 5.862  \\ 
\hline
Esophagus     & 0.649  & 12.888  & 0.677  & 10.147  \\ 
\hline
GTV     & 0.241  & 36.208  & 0.267  & 39.481  \\ 
\hline
\textbf{Overall}    &\textbf{0.744} &\textbf{22.111} &\textbf{0.756} &\textbf{12.033}\\
\hline
\end{tabular}
\bigskip
\label{tab:results}
\begin{tabular}{|l|l|l|l|l|}
\hline
   & \multicolumn{2}{l|}{\textbf{Without SS pre-training}} & \multicolumn{2}{l|}{\textbf{With SS pre-training}}  \\ 
\cline{2-5}
OH    & \textbf{Dice} & \textbf{HD95 (mm) }  & \textbf{Dice} & \textbf{HD95 (mm) }     \\ 
\hline
Lung R          & 0.971   & 9.729  & 0.973  & 20.245  \\ 
\hline
Lung L          & 0.968   & 6.275  & 0.971  & 6.416  \\ 
\hline
Spinal Cord     & 0.858  & 13.847  & 0.850  & 14.338  \\ 
\hline
Esophagus     & 0.576  & 21.056  & 0.589  & 19.739  \\ 
\hline
GTV     & 0.449  & 135.295  & 0.486  & 115.423  \\ 
\hline
\textbf{Overall}    &\textbf{0.764} &\textbf{37.240} &\textbf{0.773} &\textbf{35.232}\\
\hline
\end{tabular}
\vspace{-5mm}
\end{table}

Our results indicate that the usage of the pre-training weights improved the DICE scores and the HD95 scores.  For the relatively easier-to-segment organs (left and right lungs), we observed an improvement of 0.05\% and 0.25\% for NSCLC and Orlando Health (OH) datasets, relatively. For moderately hard-to-segment organs (spinal cord), we see an improvement of 0.6\% for the NSCLC dataset but no improvement for the OH dataset. We have achieved the best improvement when it comes to hard-to-segment organs and the small lung lesions , which we consider the hardest tissue because of the variability of their locations. For the Esophagus, which has the least volume all among other organs, we saw an improvement of 4.3\% and 2.3\% and for the lung lesions we achieved an improvement of 10.8\% and 8.2\% on NSCLC and OH datasets, respectively. Our results indicate that self-supervised pre-training brings the largest benefit to the hard-to-segment regions and organs.


In Figure \ref{fig:results}, we present example source images, the ground truth segmentation, and our network's predictions for the cases of both without and with self-supervised pre-training. At the far right, we also include a close-up heatmap of an output segmentation showing the uncertainty areas predicted for the OARs and lung lesions, and we show that uncertainty occurs only at the boundaries.  We believe this is because our MC-Swin-U model is learning at both the local and global level -- generating consistent results for the organs and lung lesions interior regions even when random dropout is enabled during model inference.


\vspace{-3mm}
\section{Conclusion}


In this study, we have shown that pre-training with public datasets can significantly improve the self-supervised segmentation of medical images. Our approach was evaluated using both public and private 3D CT datasets of lung cancer patients, and our results demonstrated improved segmentation accuracy, particularly for hard-to-segment organs and lung tumors. Our results improved by 1.6\% and 1.1\% for our public and private datasets, respectively. Our findings suggest that pre-training with public datasets can be an effective approach to enhance the performance of segmentation models, and we anticipate that as more public datasets become available, the generality of pre-trained models will improve even further.

One important aspect that we believe can enhance the usefulness of our segmentation model in clinical applications is the quantification of uncertainty in the model's segmentation outputs. This information can provide valuable insights for clinicians reviewing the model predictions, as it enables them to identify regions where the model is confident in its segmentation and areas where it is more uncertain. For our future work, we plan to continue the pursuit of uncertainty quantification to guide the model and further improve segmentation results.

Moreover, we believe that the success of our approach highlights the potential of transfer learning in medical imaging. The availability of large public datasets and pre-trained models has the potential to accelerate the development of new medical imaging applications and improve clinical decision-making. By leveraging pre-trained models, researchers and clinicians can develop more accurate and efficient models without requiring extensive labeled data.








\bibliographystyle{IEEEtran}

\bibliography{refs}

\end{document}